\documentclass[3p,times,twocolumn]{elsarticle}

\usepackage{ecrc}

\volume{00}

\firstpage{1}

\journalname{Nuclear Physics B Proceedings Supplement}

\runauth{}

\jid{nuphbp}

\jnltitlelogo{Nuclear Physics B Proceedings Supplement}

\usepackage{amssymb}

\usepackage[figuresright]{rotating}

\begin{document}

\begin{frontmatter}



\dochead{}

\title{On Branching Ratios of $B_s$ Decays and the Search for New Physics in
$B^0_s\to \mu^+\mu^-$}


\author[a,b]{Robert Fleischer}
\address[a]{Nikhef, Science Park 105, NL-1098 XG Amsterdam, Netherlands}
\address[b]{Department of Physics and Astronomy, Vrije Universiteit Amsterdam,
NL-1081 HV Amsterdam, Netherlands}

\begin{abstract}
The LHCb experiment has recently established a sizable width difference 
between the mass eigenstates of the $B_s$-meson system. This phenomenon leads 
to a subtle difference at the 10\% level between the experimental branching ratios of 
$B_s$ decays extracted from time-integrated, untagged data samples and their 
theoretical counterparts. Measuring the corresponding effective $B_s$-decay lifetimes, 
both branching ratio concepts can be converted into each other. The rare decay 
$B^0_s\to \mu^+\mu^-$ and the search for New Physics through this channel is 
also affected by this effect, which enhances the Standard-Model reference value of 
the branching ratio by ${\cal O}(10\%)$, while the effective lifetime offers a new 
observable to search for physics beyond the Standard Model that is complementary 
to the branching ratio.
\end{abstract}


\end{frontmatter}


\section{Introduction}\label{sec:intro}
Weak decays of $B^0_s$ mesons encode valuable information about the quark-flavour
sector of the Standard Model (SM) of particle physics. The conceptually simplest 
observables are branching ratios, which describe the probability for the considered 
decay to occur. 

Measurements of $B_s$-decay branching ratios at hadron colliders would require 
precise knowledge of the $B_s$ production cross section, which is not available, and 
rely therefore on experimental control channels and the ratio of the $f_s/f_{u,d}$ 
fragmentation functions (for a detailed discussion, see Ref.~\cite{FST}). At the 
$e^+e^-$ $B$ factories operating at the $\Upsilon(5S)$ resonance, $B_s$-decay 
branching ratios can be extracted since the total number of produced $B_s$ mesons 
can be determined separately \cite{Drutskoy}. 
 
The neutral $B_s$ mesons exhibit $B^0_s$--$\bar B^0_s$ mixing. Measuring the
time-dependent angular distribution of the $B^0_s \to J/\psi \phi$ decay products \cite{MPA}, 
LHCb has recently established a 
non-vanishing difference $\Delta\Gamma_s$ between the decay 
widths of the $B_s$ mass eigenstates  \cite{LHCb-DGs}:
\begin{equation}\label{ys}
	y_s \equiv \frac{\Delta\Gamma_s}{2\,\Gamma_s}\equiv
	\frac{\Gamma_{\rm L}^{(s)} - \Gamma_{\rm H}^{(s)}}{2\,\Gamma_s}= 0.088 \pm 0.014,
\end{equation}
where $\Gamma_s$ is the inverse of the average $B_s$ lifetime $\tau_{B_s}$. 
Since a discrete ambiguity could also be resolved \cite{LHCb-ambig}, we are left 
with the sign in (\ref{ys}), which agrees with the SM expectation. 
A sizable value of $\Delta\Gamma_s$ was theoretically expected since decades \cite{lenz}. 

In view of the sizable $\Delta\Gamma_s$, special care 
has to be taken when dealing with the concept of a branching ratio, and the question of 
how to convert measured ``experimental" $B_s$-decay branching ratios into 
``theoretical" $B_s$ branching ratios arises. This issue is the central topic of this 
writeup, summarizing the results of Refs.~\cite{BR-paper,Bsmumu-paper}. A special
emphasis will be put on the rare decay $B^0_s\to \mu^+\mu^-$.

\section{Branching Ratios of $B_s$ Decays}\label{sec:BR}
\subsection{Experimental vs.\ Theoretical Branching Ratios}
The untagged rate of a $B_s$ decay, were no distinction between initially 
present $B^0_s$ or $\bar B^0_s$ mesons is made, is a sum of two exponentials:
\begin{eqnarray}
	\lefteqn{\langle \Gamma(B_s(t)\to f)\rangle
	\equiv\ \Gamma(B^0_s(t)\to f)+ \Gamma(\bar B^0_s(t)\to f) }\nonumber\\
&&=R^f_{\rm H} e^{-\Gamma_{\rm H}^{(s)} t} + R^f_{\rm L} 
e^{-\Gamma_{\rm L}^{(s)} t},\label{untagged-1}
\end{eqnarray}
which can be rewritten as
\begin{eqnarray}
	\lefteqn{\langle \Gamma(B_s(t)\to f)\rangle=
	\left(R^f_{\rm H} + R^f_{\rm L}\right) e^{-\Gamma_s\,t}}\nonumber\\
&& \times	\left[ \cosh\left(y_s\, t / \tau_{B_s}\right)+
	{\cal A}^f_{\rm \Delta\Gamma}\,\sinh\left(y_s\, t/\tau_{B_s}\right)\right].\label{untagged-2}
\end{eqnarray}
Here the parameter $y_s$ was introduced in (\ref{ys}), and 
\begin{equation}
{\cal A}^f_{\rm \Delta\Gamma}\equiv\frac{R^f_{\rm H} - R^f_{\rm L}}{R^f_{\rm H} + R^f_{\rm L}}
\end{equation}
is an observable depending on the final state $f$. 
The branching ratios given by experiments are extracted from total event yields, without 
taking time information into account, and can be defined as follows \cite{BR-paper,DFN}:
\begin{equation}\label{BR-exp}
{\rm BR}\left(B_s \to f\right)_{\rm exp} 
	\equiv \frac{1}{2}\int_0^\infty \langle \Gamma(B_s(t)\to f)\rangle\, dt
\end{equation}
\begin{displaymath}	
= \frac{1}{2}\left[ \frac{R^f_{\rm H}}{\Gamma^{(s)}_{\rm H}} + 
	\frac{R^f_{\rm L}}{\Gamma^{(s)}_{\rm L}}\right]
	= \frac{\tau_{B_s}}{2}\left(R^f_{\rm H} + R^f_{\rm L}\right)
	\left[\frac{1 + {\cal A}^f_{\Delta\Gamma}\, y_s}{1-y_s^2} \right].
\end{displaymath}
On the other hand, theorists usually consider and calculate the following 
CP-averaged branching ratios:
\begin{eqnarray}
\lefteqn{{\rm BR}\left(B_s \to f\right)_{\rm theo}\equiv 
	\frac{\tau_{B_s}}{2}\langle \Gamma(B^0_s(t)\to f)\rangle\Big|_{t=0}}\nonumber\\
&& = \frac{\tau_{B_s}}{2}\left(R^f_{\rm H} + R^f_{\rm L}\right),\label{BR-theo}
\end{eqnarray}
where the $B^0_s$--$\bar B^0_s$ oscillations are ``switched off" by choosing $t=0$.
It should be noted that this $B_s$ branching ratio concept allows a straightforward comparison 
with branching ratios of decays of $B^0_d$ or $B^+_u$ mesons through the $SU(3)_{\rm F}$ 
flavour symmetry of strong interactions. 

The experimental branching ratios defined in (\ref{BR-exp}) can be converted into the  
theoretical branching ratios defined in (\ref{BR-theo}) by means of 
\begin{eqnarray}
        \lefteqn{{\rm BR}\left(B_s \to f\right)_{\rm theo}}\nonumber\\
&&   = \left[\frac{1-y_s^2}{1 + {\cal A}^f_{\Delta\Gamma}\, y_s}\right]
{\rm BR}\left(B_s \to f\right)_{\rm exp},\label{BRratio-1}
\end{eqnarray}
where the term in square brackets would equal one for a vanishing $B_s$ decay width
difference. However, for the experimental value of $y_s$ in (\ref{ys}), the theoretical $B_s\to f$ 
branching ratios can differ from their experimental counterparts by up to $10\%$, depending 
on the the final state $f$. Using theoretical input, in particular the $SU(3)_{\rm F}$ flavour
symmetry, the ${\cal A}^f_{\Delta\Gamma}$ observables can be estimated for specific 
decays (for examples of recent analyses, see Refs.~\cite{RK-lifetimes,FKR,BFK}), 
as compiled in Ref.~\cite{BR-paper}. 

\subsection{Effective $B_s$ Decay Lifetimes}
Once time information for the untagged $B_s$ decay data sample becomes available, 
the theoretical input for determining ${\cal A}_{\Delta\Gamma}$ can be avoided in the 
extraction of the theoretical branching ratio (\ref{BR-theo}) \cite{BR-paper}. 

Using the effective $B_s$ decay lifetime
\begin{eqnarray}
        \lefteqn{\tau_f \equiv \frac{\int_0^\infty t\,\langle \Gamma(B_s(t)\to f)\rangle\, dt}
        {\int_0^\infty \langle \Gamma(B_s(t)\to f)\rangle\, dt}}\nonumber \\ 
         &&\quad = \frac{\tau_{B_s}}{1-y_s^2}\left[\frac{1+2\,{\cal A}^f_{\Delta\Gamma}y_s + y_s^2}
        {1 + {\cal A}^f_{\Delta\Gamma} y_s}\right],
        \label{taueff}
\end{eqnarray}
we obtain
\begin{eqnarray}
       \lefteqn{{\rm BR}\left(B_s \to f\right)_{\rm theo}}\nonumber\\
      &&  = \left[2 - \left(1-y_s^2\right)\frac{\tau_f}{\tau_{B_s}}\right]{\rm BR}\left(B_s \to f\right)_{\rm exp},
        \label{BRratioT}
\end{eqnarray}
where only measurable quantities appear on the right-hand side. The measurement of effective 
$B_s$ decay lifetimes is hence an integral part of the extraction of the theoretical branching ratios 
(\ref{BR-theo}) from the data and not only an interesting topic to constrain the
$B^0_s$--$\bar B^0_s$ mixing parameters \cite{RK-lifetimes}. The use of the theoretically
clean relation in (\ref{BRratioT}) is advocated for the compilation of $B_s$ decay properties 
in particle listings.

For a discussion of experimental subtleties related to the measurement of $B_s$ decay 
branching ratios and effective lifetimes, the reader is referred to Ref.~\cite{BR-paper}.

\subsection{$B_s\to VV$ Decays}
The branching ratio measurements of $B_s\to VV$ decays into two vector mesons, 
such as  $B_s\to J/\psi \phi$, $B_s\to K^{*0}\bar K^{*0}$  and $B_s\to D_s^{*+}D_s^{*-}$, 
are also affected by the sizable width difference $\Delta\Gamma_s$ 
\cite{LHCb-BsKastKast,DGMV}. Here an angular analysis of the decay products of the
vector mesons has to be performed in order to disentangle the CP-even ($0, \parallel$) 
and CP-odd ($\perp$) final states, with 
\begin{equation}
	f_{VV,k}^{\rm exp}\equiv{\rm BR}^{VV,k}_{\rm exp}/{\rm BR}^{VV}_{\rm exp}.
\end{equation}
The experimental branching ratios can then be converted into the theoretical branching 
ratios through
\begin{equation}
\mbox{BR}_{\rm theo}^{VV}=\left(1-y_s^2\right) 
\Biggl[ \sum_{k=0,\parallel,\perp}  \frac{f_{VV,k}^{\rm exp}}{1+
y_s{\cal A}_{\Delta\Gamma}^{VV,k}}\Biggr]
\mbox{BR}_{\rm exp}^{VV}
\end{equation}
with the help of theoretical information on the ${\cal A}_{\Delta\Gamma}^{VV,k}$
observables, or by means of the relation
\begin{equation}
	{\rm BR}_{\rm theo}^{VV}
	= \mbox{BR}_{\rm exp}^{VV}\sum_{k=0,\parallel,\perp} 
	\biggl[2 - \left(1-y_s^2\right) \frac{\tau_k^{VV}}{\tau_{B_s}}\biggr]
	f_{VV,k}^{\rm exp}
\end{equation}
utilizing effective lifetime measurements \cite{BR-paper}.

\section{$B^0_s\to \mu^+\mu^-$ and a New Window for New Physics}\label{sec:Bsmumu}
\subsection{Setting the Stage}
A key probe of New Physics (NP) is the rare decay $B^0_s\to\mu^+\mu^-$, which receives 
only loop contributions from box and penguin topologies in the SM. Here the (theoretical) 
branching ratio is predicted as follows \cite{buras}: 
\begin{equation}\label{BR-SM}
\mbox{BR}(B_s\to \mu^+\mu^-)_{\rm SM}=(3.2\pm0.2)\times 10^{-9}.
\end{equation}
In the presence of NP, the branching ratio may be affected by new particles 
in the loops or new contributions at the tree level \cite{BG}. The error of
(\ref{BR-SM}) is dominated by lattice QCD input for non-perturbative physics \cite{gamiz}.

The limiting factor for the $B^0_s\to \mu^+\mu^-$ branching ratio measurement at
hadron colliders is the ratio $f_s/f_d$, where the fragmentation functions $f_{s(d)}$
describe the probability that a $b$ quark fragments in a $\bar B^0_{s(d)}$ meson. A new
method for determining $f_s/f_d$ using nonleptonic $\bar B^0_s\to D_s^+\pi^-$, 
$B^0_d\to D^+K^-$, $B^0_d\to D^+\pi^-$ decays \cite{FST,FST-fact} was recently 
implemented at LHCb \cite{LHCb-hadr}, with a result in good agreement with 
measurements using semileptonic decays \cite{LHCb-sl}. The $SU(3)$-breaking form-factor 
ratio entering the non-leptonic method has recently been calculated with lattice
QCD \cite{gamiz,FF-lat}.

Searches for the $B^0_s\to \mu^+\mu^-$ decay were performed by the
CDF \cite{CDF-Bsmumu}, D0  \cite{D0-Bsmumu}, ATLAS \cite{ATLAS-Bsmumu},
CMS \cite{CMS-Bsmumu} and LHCb collaborations. The latter experiment has recently
reported the currently most stringent upper bound on the branching ratio, which corresponds to 
\begin{equation}\label{LHCb-bound}
\mbox{BR}(B_s\to \mu^+\mu^-)< 4.5 \times 10^{-9}
\end{equation}
at the 95\% confidence level  \cite{LHCb-Bsmumu}, and is approaching the SM prediction
(\ref{BR-SM}). For a recent review of the experimental studies, see Ref.~\cite{albrecht}.

In the analyses of the $B^0_s\to \mu^+\mu^-$ decay, the impact of $\Delta\Gamma_s$ was
not taken into account. In view of the discussion in the previous section, the question arises 
how the sizable value of $\Delta\Gamma_s$ affects the theoretical interpretation of the 
$B^0_s\to \mu^+\mu^-$ data and whether we can actually take advantage of this 
decay width difference \cite{Bsmumu-paper}. 

\subsection{Low-Energy Effective Hamiltonian}
In order to address this issue, we use the general low-energy effective Hamiltonian 
describing the decay $\bar B^0_s\to \mu^+\mu^-$ as the starting point. Using a notation
similar to Ref.~\cite{APS}, where a model-independent NP analysis was performed,
it can be  written as
\begin{eqnarray}
\lefteqn{{\cal H}_{\rm eff}=-\frac{G_{\rm F}}{\sqrt{2}\pi} \alpha V_{ts}^\ast V_{tb} 
\bigl[C_{10} O_{10} + C_{S} O_S + C_P O_P}\nonumber\\
&& + C_{10}' O_{10}' + C_{S}' O_S' + C_P' O_P' \bigr],\label{Heff}
\end{eqnarray}
where $G_{\rm F}$ and $\alpha$ are the Fermi and QED fine-structure constants,
respectively, and the $V_{qq'}$ are elements of the Cabibbo--Kobayashi--Maskawa (CKM) matrix.
The short-distance physics is encoded in the Wilson coefficients $C_i$,  $C_i'$ of the
four-fermion operators
\begin{eqnarray}
O_{10}&=&(\bar s \gamma_\mu P_L b) (\bar\ell\gamma^\mu \gamma_5\ell) \nonumber\\
O_S&=&m_b (\bar s P_R b)(\bar \ell \ell) \\
O_P&=&m_b (\bar s P_R b)(\bar \ell \gamma_5 \ell), \nonumber
\end{eqnarray}
where $m_b$ denotes the $b$-quark mass, $P_{L,R}\equiv(1\mp\gamma_5)/2$, and the
$O'_i$  are obtained from the $O_i$ through the replacements $P_L \leftrightarrow P_R$.
It should be noted that only operators with non-vanishing contributions to 
$\bar B^0_s\to \mu^+\mu^-$ are included in (\ref{Heff}); in particular the matrix elements
of operators involving the $\bar \ell\gamma^\mu\ell$ vector current vanish.

The hadronic structure of the leptonic $\bar B^0_s\to \mu^+\mu^-$ decay is very simple
and can be expressed in terms of a single, non-perturbative parameter, which is the 
$B_s$-meson decay constant $f_{B_s}$ \cite{buras}.

In the SM, only the $O_{10}$ operator contributes with a real Wilson coefficient $C_{10}^{\rm SM}$,
leading to the prediction in (\ref{BR-SM}). The sensitivity to (pseudo-)scalar lepton densities
entering the $O_{(P)S}$ and $O_{(P)S}'$ operators is an outstanding feature of the  
$\bar B^0_s\to \mu^+\mu^-$ channel. The corresponding Wilson coefficients are still 
largely unconstrained, thereby leaving ample space for  NP \cite{APS}.

\subsection{Observables}
For the calculation of the $B_s\to \mu^+\mu^-$ observables, it is convenient to go to
the rest frame of the decaying $\bar B^0_s$ meson and to use the notation
$\mu_\lambda^+\mu_\lambda^-$ to distinguish between the left-handed ($\lambda={\rm L}$) 
and right-handed ($\lambda={\rm R}$) muon helicity configurations. In this setting, the 
$\mu^+_{\rm L}\mu^-_{\rm L}$ and $\mu^+_{\rm R}\mu^-_{\rm R}$ states are simply related 
to each other through CP transformations.

Thanks to $B^0_s$--$\bar B^0_s$ mixing, we get interference effects between the
$\bar B^0_s\to \mu_\lambda^+\mu_\lambda^-$ and $B^0_s\to \mu_\lambda^+\mu_\lambda^-$
decay processes that are described by the observable
\begin{equation}\label{xi-def}
\xi_\lambda\equiv - e^{-i\phi_s}\left[ e^{i\phi_{\rm CP}(B_s)}
\frac{A(\bar B^0_s\to \mu_\lambda^+\mu_\lambda^-)}{A(B^0_s\to \mu_\lambda^+\mu_\lambda^-)}
\right].
\end{equation}
Here $\phi_s$ is the $B^0_s$--$\bar B^0_s$ mixing phase, whereas 
$\phi_{\rm CP}(B_s)$ denotes a convention-dependent phase which is associated with
CP transformations \cite{RF-habil}. Expressing the 
$\bar B^0_s \to \mu_\lambda^+\mu_\lambda^-$ decay amplitude as
\begin{equation}\label{ampl}
A(\bar B^0_s \to \mu_\lambda^+\mu_\lambda^-)=\langle \mu_\lambda^-\mu_\lambda^+|
{\cal H}_{\rm eff}| \bar B^0_s \rangle
\end{equation}
and performing appropriate CP transformation results eventually in the following expression 
\cite{Bsmumu-paper}:
\begin{equation}\label{xi-obs}
\xi_\lambda=-\left[\frac{+\eta_\lambda P \,+\,  S}{-\eta_\lambda P^\ast + S^\ast }\right],
\end{equation}
where all convention-dependent quantities (such as the $\phi_{\rm CP}(B_s)$ phases) 
cancel, $\eta_{\rm L(R)}=+(-)1$, and
\begin{equation}\label{P-expr}
P\equiv \frac{C_{10}-C_{10}'}{C_{10}^{\rm SM}}+\frac{M_{B_s} ^2}{2 m_\mu}
\left(\frac{m_b}{m_b+m_s}\right)\left(\frac{C_P-C_P'}{C_{10}^{\rm SM}}\right)
\end{equation}
\begin{equation}\label{S-expr}
S\equiv \sqrt{1-4\frac{m_\mu^2}{M_{B_s}^2}}
\frac{M_{B_s} ^2}{2 m_\mu}\left(\frac{m_b}{m_b+m_s}\right)
\left(\frac{C_S-C_S'}{C_{10}^{\rm SM}}\right).
\end{equation}
These combinations of Wilson coefficient functions have been introduced in such a way 
that we simply have $P=1$ and $S=0$ in the SM, whereas $P\equiv |P|e^{i\varphi_P}$ and 
$S\equiv |S|e^{i\varphi_S}$ carry, in general, non-trivial CP-violating phases $\varphi_P$ 
and $\varphi_S$ (see also Ref.~\cite{APS}). It should be noted that the non-perturbative
$B_s$-meson decay constant $f_{B_s} $, which arises in the parametrization of (\ref{ampl}) 
and affects the SM prediction (\ref{BR-SM}), cancels in the observable (\ref{xi-obs}). 

Before having a closer look at the branching ratio, it is interesting to consider
the following time-dependent rate asymmetries, which require tagging information
and knowledge of the muon helicity $\lambda$:
\begin{displaymath}
\frac{\Gamma(B^0_s(t)\to \mu_\lambda^+\mu^-_\lambda)-
\Gamma(\bar B^0_s(t)\to \mu_\lambda^+
\mu^-_\lambda)}{\Gamma(B^0_s(t)\to \mu_\lambda^+\mu^-_\lambda)+
\Gamma(\bar B^0_s(t)\to \mu_\lambda^+\mu^-_\lambda)}
\end{displaymath}
\vspace*{-0.3truecm}
\begin{equation}\label{asym-1}
=\frac{C_\lambda\cos(\Delta M_st)+S_\lambda\sin(\Delta M_st)}{\cosh(y_st/\tau_{B_s}) + 
{\cal A}_{\Delta\Gamma}^\lambda \sinh(y_st/\tau_{B_s})}.
\end{equation}
Here $\Delta M_s$ denotes the mass difference between the heavy and light 
$B_s$ mass eigenstates while $y_s$ is given in (\ref{ys}). Neglecting the impact
of $\Delta\Gamma_s$, such CP asymmetries were considered for $B_{s,d}\to\ell^+\ell^-$ 
decays within various NP scenarios in the previous literature \cite{HL,DP,CKWW}.

The observables entering (\ref{asym-1}) are governed by $\xi_\lambda$ in (\ref{xi-obs})
and take the following expressions \cite{Bsmumu-paper}:
\begin{equation}\label{C-lam}
\hspace*{-0.2truecm}C_\lambda\equiv\frac{1-|\xi_\lambda|^2}{1+|\xi_\lambda|^2}
=-\eta_\lambda\left[\frac{2|PS|\cos(\varphi_P-\varphi_S)}{|P|^2+|S|^2}  \right]
\end{equation}
\begin{equation}\label{S-lam}
\hspace*{-0.2truecm}S_\lambda\equiv \frac{2\,\mbox{Im}\,\xi_\lambda}{1+|\xi_\lambda|^2}
=\frac{|P|^2\sin 2\varphi_P-|S|^2\sin 2\varphi_S}{|P|^2+|S|^2}
\end{equation}
\begin{equation}\label{ADG-lam}
\hspace*{-0.2truecm}
{\cal A}_{\Delta\Gamma}^\lambda\equiv \frac{2\,\mbox{Re}\,\xi_\lambda}{1+|\xi_\lambda|^2}
=\frac{|P|^2\cos 2\varphi_P-|S|^2\cos 2\varphi_S}{|P|^2+|S|^2},
\end{equation}
which are theoretically clean. Note that ${\cal S}_{\rm CP} \equiv S_\lambda$ and 
${\cal A}_{\Delta\Gamma}\equiv {\cal A}_{\Delta\Gamma}^\lambda$ do not depend on the
muon helicity $\lambda$.

In the discussion given above, it was assumed that NP enters only through the Wilson 
coefficients governing the $\bar B^0_s\to \mu^+\mu^-$ decay and that the $B^0_s$--$\bar B^0_s$ 
mixing phase $\phi_s=\phi_s^{\rm SM}+\phi_s^{\rm NP}$ takes its SM value 
$\phi_s^{\rm SM}\equiv2\mbox{arg}(V_{ts}^\ast V_{tb})$, which is cancelled in 
(\ref{xi-def}) through the CKM factors of the ratio of decay amplitudes. In (\ref{S-lam}) and 
(\ref{ADG-lam}), NP in $B^0_s$--$\bar B^0_s$ mixing can straightforwardly be included 
through the replacements $2\varphi_{P,S}\to 2\varphi_{P,S}-\phi_s^{\rm NP}$. The LHCb 
data for CP violation in $B_s\to J/\psi \phi, J/\psi f_0(980)$ decays already constrain 
$\phi_s^{\rm NP}$ to the few-degree level \cite{MPA}. Consequently, this effect is
negligible from the practical point of view for the following considerations.

It is difficult to measure the muon helicity. In order to circumvent this problem, we consider the
\begin{equation}
\Gamma({B}_s^0(t)\to \mu^+\mu^-)\equiv \sum_{\lambda={\rm L,R}}
\Gamma({B}_s^0(t)\to \mu^+_\lambda \mu^-_\lambda)
\end{equation}
rate and its counterpart for initially present $\bar B^0_s$ mesons, which
can be combined into the CP asymmetry
\begin{eqnarray}
\lefteqn{\frac{\Gamma(B^0_s(t)\to \mu^+\mu^-)-
\Gamma(\bar B^0_s(t)\to \mu^+\mu^-)}{\Gamma(B^0_s(t)\to \mu^+\mu^-)+
\Gamma(\bar B^0_s(t)\to \mu^+\mu^-)}}\nonumber\\
&&=\frac{{\cal S}_{\rm CP}\sin(\Delta M_st)}{\cosh(y_st/ \tau_{B_s}) + 
{\cal A}_{\Delta\Gamma} \sinh(y_st/ \tau_{B_s})}.
\end{eqnarray}
Since a non-zero value would immediately signal new CP-violating phases, it
would be most interesting to measure this asymmetry. This feature was recently
highlighted in Ref.~\cite{BG-2} for minimal $U(2)^3$ models \cite{barbieri}. 
Unfortunately, despite the independence on the muon helicity, this is still 
challenging from the practical point of view as tagging and time information are 
required. An analogous expression holds for the rare $B_d\to\mu^+\mu^-$ decays, 
where $y_d$ is negligibly small.

\subsection{Closer Look at the Branching Ratios}
From the practical point of view, the branching ratio extracted from untagged
data samples, ignoring the decay-time information, is the first measurement:
\begin{eqnarray}\label{defBrExp}
      \lefteqn{{\rm BR}\left(B_s \to \mu^+\mu^-\right)_{\rm exp}}\nonumber\\
       && \equiv \frac{1}{2}\int_0^\infty \langle \Gamma(B_s(t)\to \mu^+\mu^-)\rangle\, dt.
\end{eqnarray}
Here the untagged $\langle \Gamma(B_s(t)\to \mu^+\mu^-)\rangle$ rate is given 
in general terms in (\ref{untagged-1}) and (\ref{untagged-2}).

Since ${\cal A}_{\Delta\Gamma}^\lambda$ in (\ref{ADG-lam}) does actually not depend on the muon
helicity, i.e.\  ${\cal A}_{\Delta\Gamma}\equiv {\cal A}_{\Delta\Gamma}^\lambda$, we
can apply (\ref{BRratio-1}) to extract the theoretical branching ratio from the 
experimental branching ratio (\ref{defBrExp}):
\begin{eqnarray}
        \lefteqn{{\rm BR}(B_s \to \mu^+\mu^-)_{\rm theo}}\nonumber\\
        &&=   \left[\frac{1-y_s^2}{1 + {\cal A}_{\Delta\Gamma}\, y_s}\right] 
        {\rm BR}(B_s \to \mu^+\mu^-)_{\rm exp}.\label{BRratio}
\end{eqnarray}
The former is considered and calculated by the theoretical community 
(see, e.g., Refs.~\cite{buras,APS}), and satisfies 
\begin{equation}
\frac{\mbox{BR}(B_s\to\mu^+\mu^-)_{\rm theo}}{\mbox{BR}(B_s\to\mu^+\mu^-)_{\rm SM}}=
|P|^2+|S|^2.
\end{equation}
The $y_s$ terms in (\ref{BRratio}) had not been taken into account in the comparison 
between theory and experiment \cite{Bsmumu-paper}. 

As can be seen in (\ref{ADG-lam}), ${\cal A}_{\Delta\Gamma}$ depends sensitively on NP 
entering through the Wilson coefficients which govern the $B^0_s\to \mu^+\mu^-$ channel.
Consequently, this observable is currently unknown. Varying ${\cal A}_{\Delta\Gamma}\in[-1,+1]$
yields
\begin{equation}\label{BR-error}
\hspace*{-0.2truecm} \Delta {\rm BR}(B_s \to \mu^+\mu^-)|_{y_s}=\pm y_s 
{\rm BR}(B_s \to \mu^+\mu^-)_{\rm exp},
\end{equation}
which has to be added to the experimental error of (\ref{defBrExp}).

On the other hand, within the SM, we have the theoretically clean prediction 
${\cal A}_{\Delta\Gamma}^{\rm SM}=+1$. If we rescale the theoretical SM
branching ratio in (\ref{BR-SM}) correspondingly by a factor of $1/(1-y_s)$
and use (\ref{ys}), we obtain
\begin{equation}
\mbox{BR}(B_s\to \mu^+\mu^-)_{\rm SM}|_{y_s}=(3.5\pm0.2)\times 10^{-9}.
\end{equation}
This is the SM branching ratio reference value for the comparison with 
the experimental branching ratio (\ref{defBrExp}). 

\subsection{Effective Lifetime}
Once the $B^0_s\to\mu^+\mu^-$ decay has been observed and more experimental
data become available, it is possible to include also the decay time information in the 
analysis so that the effective $B^0_s\to\mu^+\mu^-$ lifetime $\tau_{\mu^+\mu^-}$,
which is defined as in (\ref{taueff}), 
can be measured. Since ${\cal A}_{\Delta\Gamma}$ in (\ref{ADG-lam}) does not depend
on the muon helicity, this observable can be extracted from the effective lifetime with
the help of the relation
\begin{equation}
 {\cal A}_{\Delta\Gamma}  = \frac{1}{y_s}\left[\frac{(1-y_s^2)\tau_{\mu^+\mu^-}-(1+
 y_s^2)\tau_{B_s}}{2\tau_{B_s}-(1-y_s^2)\tau_{\mu^+\mu^-}}\right],
\end{equation} 
and results in
\begin{eqnarray}
\lefteqn{{\rm BR}\left(B_s \to \mu^+\mu^-\right)_{\rm theo}}\nonumber\\
&&\hspace*{-0.4truecm}=\left[2 - \left(1-y_s^2\right)\frac{\tau_{\mu^+\mu^-}}{\tau_{B_s}}\right] 
{\rm BR}\left(B_s \to \mu^+\mu^-\right)_{\rm exp}.\label{BRmumu-correct}
\end{eqnarray}
This expression takes the same form as (\ref{BRratioT}) and allows 
the conversion of the experimental $B_s\to\mu^+\mu^-$ branching ratio into its theoretical 
counterpart, irrespective of whether there are NP contributions present or not. Consequently, 
the error in (\ref{BR-error}) can then be eliminated.

The effective $B^0_s\to\mu^+\mu^-$ lifetime and the extraction of ${\cal A}_{\Delta\Gamma}$
from untagged data samples is an important new measurement for the high-luminosity
upgrade of the LHC. Extrapolating from the currently available analyses of the effective
$B_s^0\to J/\psi\,f_0(980)$ and $B_s^0\to K^+K^-$ lifetimes performed by the CDF and 
LHCb collaborations, a precision of $5\%$ or better appears feasible \cite{Bsmumu-paper}. 
Detailed experimental studies of this exciting new feature of the $B^0_s\to\mu^+\mu^-$ channel
are strongly encouraged.

\subsection{Constraints on New Physics}
Looking at the expression for ${\cal A}_{\Delta\Gamma} $ in (\ref{ADG-lam}), it is obvious
that this observable and the effective lifetime $\tau_{\mu^+\mu^-}$ may well be affected
by NP. The $\Delta\Gamma_s$ effects propagate also into the constraints on NP parameters 
that can be obtained from the comparison of the experimental information on the 
$B_s\to\mu^+\mu^-$ branching ratio with the SM branching ratio, where it is 
useful to introduce  \cite{Bsmumu-paper}
\begin{equation}
R\equiv 
\frac{\mbox{BR}(B_s \to \mu^+\mu^-)_{\rm exp}}{\mbox{BR}(B_s \to \mu^+\mu^-)_{\rm SM}}.
\end{equation}\label{R-def}
Using (\ref{ADG-lam}) and (\ref{BRratio}), the ratio $R$ takes the form
\begin{eqnarray}
\lefteqn{\hspace*{-0.3truecm}R=\left[\frac{1 + {\cal A}_{\Delta\Gamma} y_s}{1-y_s^2} \right]
\left( |P|^2+ |S|^2\right)}\nonumber\\
&&\hspace*{-0.9truecm}=\left[\frac{1+y_s\cos2\varphi_P}{1-y_s^2}  \right] |P|^2+
\left[\frac{1-y_s\cos2\varphi_S}{1-y_s^2}  \right] |S|^2.\label{R-expr}
\end{eqnarray}
Combining (\ref{BR-SM}) and (\ref{LHCb-bound}) yields the bound $R<1.4$, where
the theoretical uncertainty of the SM prediction of the $B^0_s\to \mu^+\mu^-$ branching 
ratio was neglected. 

The ratio $R$ would fix a circle in the $|P|$--$|S|$ plane for $y_s=0$, i.e.\ for a 
vanishing $B_s$ decay width difference. On the other hand, for non-zero values of $y_s$, 
$R$ can be converted into ellipses which depend on the CP-violating phases $\varphi_{P,S}$.
As the latter quantities are in general unknown, $R$ results in a circular band, with the
upper bounds $|P|, |S|\leq\sqrt{(1+y_s)R}$.
Since the $S$ and $P$ contributions cannot be separated through experimental 
information on $R$, as can be seen in (\ref{R-expr}),
there may still significant NP contributions be present in 
$B^0_s\to\mu^+\mu^-$, even if the branching ratio should eventually be measured 
close to the SM expectation. 

As was pointed out in Ref.~\cite{Bsmumu-paper}, the measurement of the
effective lifetime $\tau_{\mu^+\mu^-}$ and the associated untagged 
${\cal A}_{\Delta\Gamma}$ observable allows a resolution of this situation. The point is that 
\begin{equation}\label{S-ADG}
|S|=|P|\sqrt{\frac{\cos2\varphi_P-{\cal A}_{\Delta\Gamma}}{\cos2\varphi_S
+{\cal A}_{\Delta\Gamma}}}
\end{equation}
fixes a straight line through the origin in the $|P|$--$|S|$ plane. For illustrations, the
reader is referred to the figures shown in Ref.~\cite{Bsmumu-paper}.

In the most recent analyses of the constraints on NP parameter space that are implied 
by the experimental upper bound on the $B_s\to\mu^+\mu^-$ branching ratio for various 
extensions of the SM, authors have now started to take the effect of $\Delta\Gamma_s$ 
into account (see, for instance, the papers listed in Refs.~\cite{BG-2,NP}).

\section{Conclusions}\label{sec:concl}
The non-vanishing width difference of the $B_s$-meson system, which 
has recently been established by LHCb,
leads to subtleties in the extraction of $B_s$ branching ratio information from the data
but offers also new observables. The differences between the experimental and
theoretical branching ratios can be as large as $10\%$, depending on the
final state. Both branching ratios can be converted into each other either through
theoretical considerations or  through the measurement of the effective $B_s\to f$ 
decay lifetimes. As the latter involves only experimental data, it is generally the preferred avenue.

The rare decay $B^0_s\to\mu^+\mu^-$ is also affected by $\Delta\Gamma_s$, where
the theoretical branching ratio in (\ref{BR-SM}) has to be rescaled by $1/(1-y_s)$ for 
the comparison with the experimental branching ratio, resulting in the SM reference value 
of $(3.5\pm0.2)\times 10^{-9}$. 
Thanks to $\Delta\Gamma_s$, the $B^0_s\to \mu^+\mu^-$ decay offers a new observable,
which is the effective lifetime $\tau_{\mu^+\mu^-}$. It allows the inclusion of the
$\Delta\Gamma_s$ effects in the conversion of the experimental into the theoretical
branching ratio. Moreover, it offers also a new, theoretically clean NP probe that may still 
show large NP effects, in particular those originating from the (pseudo-)scalar $\ell^+\ell^-$
densities entering the four-fermion operators. This observable may even show NP should
the $B^0_s\to \mu^+\mu^-$ branching ratio be found close to the SM prediction.
The measurement of $\tau_{\mu^+\mu^-}$ and the associated ${\cal A}_{\Delta\Gamma}$
observable is an exciting new topic for the high-luminosity upgrade of the LHC.
Detailed feasibility studies are strongly encouraged.





\begin{thebibliography}{00}
 
\bibitem{FST}R.~Fleischer, N.~Serra and N.~Tuning,
  Phys.\ Rev.\ D {\bf 82} (2010) 034038
  [arXiv:1004.3982 [hep-ph]].
  
\bibitem{Drutskoy}
  A.~Drutskoy {\it et al.}  (Belle Collaboration),
  Phys.\ Rev.\ D {\bf 76} (2007)  012002  
  [hep-ex/0610003].
  
\bibitem{MPA}M. Pepe Altarelli, these proceedings.
  
\bibitem{LHCb-DGs}R. Aaij {\it et al.}\ [LHCb Collaboration],  LHCb-CONF-2012-002.

\bibitem{LHCb-ambig}R.~Aaij {\it et al.}  [LHCb Collaboration],
  Phys.\ Rev.\ Lett.\  {\bf 108} (2012) 241801
  [arXiv:1202.4717 [hep-ex]].

\bibitem{lenz}For a recent review, see A.~Lenz,
  arXiv:1205.1444 [hep-ph].

\bibitem{BR-paper}K.~De Bruyn, R.~Fleischer, R.~Knegjens, P.~Koppenburg, M.~Merk 
and N.~Tuning,
  Phys.\ Rev.\ D {\bf 86} (2012) 014027
  [arXiv:1204.1735 [hep-ph]].
  
 \bibitem{Bsmumu-paper}K.~De Bruyn, R.~Fleischer, R.~Knegjens, P.~Koppenburg, 
 M.~Merk, A.~Pellegrino and N.~Tuning,
  Phys.\ Rev.\ Lett.\  {\bf 109} (2012) 041801
  [arXiv:1204.1737 [hep-ph]].
  
\bibitem{DFN}I.~Dunietz, R.~Fleischer and U.~Nierste,
  Phys.\ Rev.\ D {\bf 63} (2001) 114015
  [hep-ph/0012219].
  
\bibitem{RK-lifetimes}R.~Fleischer and R.~Knegjens,
  Eur.\ Phys.\ J.\ C {\bf 71} (2011) 1789
  [arXiv:1109.5115 [hep-ph]]; R. Knegjens, these proceedings.
  
\bibitem{FKR}R.~Fleischer, R.~Knegjens and G.~Ricciardi,
  Eur.\ Phys.\ J.\ C {\bf 71} (2011) 1832
  [arXiv:1109.1112 [hep-ph]];
  Eur.\ Phys.\ J.\ C {\bf 71} (2011) 1798
  [arXiv:1110.5490 [hep-ph]].
  
\bibitem{BFK}K.~De Bruyn, R.~Fleischer and P.~Koppenburg,
  Eur.\ Phys.\ J.\ C {\bf 70} (2010) 1025
  [arXiv:1010.0089 [hep-ph]].

\bibitem{LHCb-BsKastKast}R.~Aaij {\it et al.}  [LHCb Collaboration],
  Phys.\ Lett.\ B {\bf 709} (2012) 50
  [arXiv:1111.4183 [hep-ex]].

\bibitem{DGMV}S.~Descotes-Genon, J.~Matias and J.~Virto,
  Phys.\ Rev.\ D {\bf 85} (2012) 034010
  [arXiv:1111.4882 [hep-ph]].
  
\bibitem{buras}A.~J.~Buras,
  PoS BEAUTY {\bf 2011}, 008 (2011)
  [arXiv:1106.0998 [hep-ph]].

\bibitem{BG}For a recent review, see A.~J.~Buras and J.~Girrbach,
  arXiv:1204.5064 [hep-ph].

\bibitem{gamiz}E. Gamiz, these proceedings.

\bibitem{FST-fact}R.~Fleischer, N.~Serra and N.~Tuning,
  Phys.\ Rev.\ D {\bf 83} (2011) 014017
  [arXiv:1012.2784 [hep-ph]].
  
\bibitem{LHCb-hadr}R.~Aaij {\it et al.}  [LHCb Collaboration],
  Phys.\ Rev.\ Lett.\  {\bf 107} (2011) 211801
  [arXiv:1106.4435 [hep-ex]].

\bibitem{LHCb-sl}R.~Aaij {\it et al.}  [LHCb Collaboration],
  Phys.\ Rev.\ D {\bf 85} (2012) 032008
  [arXiv:1111.2357 [hep-ex]].

\bibitem{FF-lat}J.~A.~Bailey, A.~Bazavov, C.~Bernard, C.~M.~Bouchard, C.~DeTar, D.~Du, A.~X.~El-Khadra and J.~Foley {\it et al.},
  Phys.\ Rev.\ D {\bf 85} (2012) 114502
  [arXiv:1202.6346 [hep-lat]].

\bibitem{CDF-Bsmumu} 
  T.~Aaltonen {\it et al.}\  [CDF Collaboration],
  Phys.\ Rev.\ Lett.\  {\bf 107} (2011) 191801 
  [arXiv:1107.2304 [hep-ex]].

\bibitem{D0-Bsmumu}
  V.~M.~Abazov {\it et al.}\  [D0 Collaboration],
  Phys.\ Lett.\ B {\bf 693} (2010) 539 
  [arXiv:1006.3469 [hep-ex]].
  
\bibitem{ATLAS-Bsmumu} 
G.~Aad {\it et al.}\  [ATLAS Collaboration],
  Phys.\ Lett.\ B {\bf 713} (2012) 387
  [arXiv:1204.0735 [hep-ex]]; M. Bona, these proceedings.
  
  \bibitem{CMS-Bsmumu} 
 S.~Chatrchyan {\it et al.}\  [CMS Collaboration],
  JHEP {\bf 1204} (2012) 033
  [arXiv:1203.3976 [hep-ex]]; G. Tonelli, these proceedings.

\bibitem{LHCb-Bsmumu}R.~Aaij {\it et al.}\  [LHCb Collaboration],
  Phys.\ Rev.\ Lett.\ {\bf 108} (2012) 231801 [arXiv:1203.4493 [hep-ex]]; J. Albrecht, 
  these proceedings.

\bibitem{albrecht}J.~Albrecht,
  arXiv:1207.4287 [hep-ex].
  
\bibitem{APS}W.~Altmannshofer, P.~Paradisi and D.~M.~Straub,
  JHEP {\bf 1204} (2012) 008
  [arXiv:1111.1257 [hep-ph]].
  
  \bibitem{RF-habil}R.~Fleischer,
  Phys.\ Rept.\  {\bf 370} (2002) 537  [hep-ph/0207108].

\bibitem{HL} C.-S.~Huang and W.~Liao,
  Phys.\ Lett.\ B {\bf 525} (2002) 107  [hep-ph/0011089];
  Phys.\ Lett.\ B {\bf 538} (2002) 301
  [hep-ph/0201121].
  
\bibitem{DP}A.~Dedes and A.~Pilaftsis,
  Phys.\ Rev.\ D {\bf 67} (2003) 015012  [hep-ph/0209306].

\bibitem{CKWW}P.~H.~Chankowski, J.~Kalinowski, Z.~Was and M.~Worek,
  Nucl.\ Phys.\ B {\bf 713} (2005) 555  [hep-ph/0412253].
    
\bibitem{BG-2}A.~J.~Buras and J.~Girrbach,
  arXiv:1206.3878 [hep-ph].

\bibitem{barbieri}R. Barbieri, these proceedings.
    
\bibitem{NP}O.~Buchmueller, R.~Cavanaugh, M.~Citron, A.~De Roeck, M.~J.~Dolan, J.~R.~Ellis, H.~Fl\"acher and S.~Heinemeyer {\it et al.},
  arXiv:1207.7315 [hep-ph];
   T.~Hurth and F.~Mahmoudi,
  arXiv:1207.0688 [hep-ph];
   W.~Altmannshofer and D.~M.~Straub,
  arXiv:1206.0273 [hep-ph];
  D.~Becirevic, N.~Kosnik, F.~Mescia and E.~Schneider,
  arXiv:1205.5811 [hep-ph];
  F.~Mahmoudi, S.~Neshatpour and J.~Orloff,
  arXiv:1205.1845 [hep-ph];
   T.~Li, D.~V.~Nanopoulos, W.~Wang, X.~-C.~Wang and Z.~-H.~Xiong,
  JHEP {\bf 1207} (2012) 190
  [arXiv:1204.5326 [hep-ph]].
  
 \end{thebibliography}



\section*{Acknowledgements}
I am grateful to the organizers, in particular Giulia Ricciardi, for hosting and inviting me to
another excellent meeting of the Capri flavour physics workshop series, and would like to 
thank my PhD students and colleagues for the enjoyable collaboration on the topics
discussed above.

\end{document}